\documentstyle[12pt]{article}
\begin{document}
\title{Probabilistic Super Dense Coding}
%\centerline{\Large \bf Probabilistic Super Dense Coding}
%\vspace{4ex}
%\begin{center}
\author{A. K. Pati$^{1}$, P. Parashar$^{2}$ and P. Agrawal$^{1}$ \\
$^{1}$Institute of Physics, Sainik School Post,\\
Bhubaneswar-751005, India \\                                
$^{2}$ Physics and Applied Mathematics Unit,\\
Indian Statistical Institute, Kolkata-700108, India }

%\end{center}

\newcommand{\Nol}{{1 \over \sqrt{1 + |\ell|^ 2} }}
\newcommand{\Non}{{1 \over \sqrt{1 + |n|^2    } }}
\newcommand{\Nop}{{1 \over \sqrt{1 + |p|^2    } }}
\newcommand{\Nolp}{{1 \over \sqrt{1 + |\ell^{\prime}|^ 2} }}
\newcommand{\Nom}{{1 \over \sqrt{1 + |m|^2    } }}
\newcommand{\Nopp}{{1 \over \sqrt{1 + |p^{\prime}|^2    } }}

\newcommand{\be}{\begin{equation}}
\newcommand{\ee}{\end{equation}}
\newcommand{\bea}{\begin{eqnarray}}
\newcommand{\eea}{\end{eqnarray}}

\newcommand{\rag}{\rangle}
\def\ie{\hbox{\it i.e.}{}}
\def\eg{\hbox{\it e.g.}{}}

%\date{\today}
\maketitle

\def\ra{\rangle}
\def\la{\langle}
\def\ver{\arrowvert}

\vspace{10ex}

\begin{abstract}
%It is known that one can double the classical communication capacity 
%by sending a qubit which is part of an apriory shared  maximally entangled 
%state. However, with a non-maximally entangled state one cannot perform 
%super dense coding in a perfect fashion. The resulting classical information
%will be noisy. 
We explore the possibility of performing super dense coding with 
non-maximally entangled states as a resource. Using this we find that one can 
send two classical bits in a 
probabilistic manner by sending a qubit. We generalize our scheme to 
higher dimensions and show that one can communicate $2\log_2 d$ classical 
bits by sending a $d$-dimensional quantum state with a certain probability 
of success. The success probability in super dense coding is related to 
the success probability of distinguishing non-orthogonal states. 
The optimal average success probabilities are explicitly calculated.
We consider the possibility of sending $2 \log_2 d$ classical bits with 
a shared resource of a 
higher dimensional entangled state $( D \times D, D > d)$. It is
found that more entanglement does not necessarily lead to higher success 
probability. This also answers the question as to why  we need $\log_2 d$ 
ebits to send $2\log_2 d$ classical bits in a deterministic fashion.

\end{abstract}

%\vfill

%\pagestyle{empty}

\vskip 1cm

PACS           NO:    03.67.-a, 03.65.Bz\\

\vskip 2cm

%\footnote{
%\noindent
%E-mail:\\

$\overline{{\rm Emails: 
akpati@iopb.res.in, parashar@isical.ac.in, agrawal@iopb.res.in }}$

%\begin{multicols}{2}

\newpage

\par

\section{Introduction }
It is by now, well demonstrated that entangled states are at the heart of 
quantum
information theory. One can do many surprising tasks using entangled states
which are otherwise impossible, e.g., super dense coding \cite{bw}, quantum 
teleportation \cite{bbc}, remote state preparation \cite{akp}, 
quantum cryptography \cite{grtz} and
so on. In the case of super dense coding, Bennett and Wisner have 
shown that it is possible
to send two classical bits of information by sending just a single qubit
\cite{bw}. Ordinarily by sending a single qubit one would extract only 
one bit of classical information. However, 
prior sharing of entangled state enhances the classical 
communication capacity, hence the name super dense coding. In a similar 
fashion, if one
shares $\log_2 d$ ebits of entanglement then one can extract $2\log_2 d$
classical bits of information by sending a $d$-level quantum system
(a qudit). 

In recent years, super dense coding has been generalized in various
directions. For example, it is possible to 
generalize the super dense coding for multi-parties \cite{bose}.
Also, one can perform super dense coding not only with  quantum states 
in finite dimensional Hilbert spaces but
also with quantum states in infinite dimensional Hilbert spaces 
\cite{bk,bp}. All these cases deal with 
maximally entangled (ME) states.
But suppose Alice and Bob share a non-maximally entangled (NME) state, 
then what can they do? This question was first addressed by Barenco and 
Ekert \cite{be}. However, their scheme is not a conclusive one. 
It was shown by Hausladen {\it et al} \cite{haus} 
that if one has a 
non-maximally entangled state then the classical capacity of dense coding 
scheme is not $2\log_2 d$
but equal to $H_E ~+~ \log_2 d$ bits of information in the asymptotic limit, 
where $H_E$ is the 
entropy of entanglement of the shared state. Here, $0 \le H_E \le \log_2 d$.
However, the above scheme is a deterministic one. So this result tells us that
deterministically we cannot send $2\log_2 d$ bits using NME states. 
The super dense coding protocol has been generalized for mixed entangled states
and the classical capacity has been related to various measures of entanglement
\cite{bpv}.
Very recently,
Mozes {\it et al} \cite{mozes} have investigated the relationship between the
entanglement of a given NME state and the maximum number of alphabets which
can be perfectly transmitted in a deterministic fashion (this is called `not
so super dense coding').
%shown that for Hilbert space of 
%dimension
%$d > 2$ it is possible to have deterministic dense coding by consuming less
%than one ebit. Moreover, they have shown that it is not possible to have 
%deterministic dense coding in Hilbert space of dimension $d < 3$.

All the previous work are primarily on deterministic super dense
coding.  If one does not demand that the scheme works in a
deterministic manner, then it should be possible to send $2\log_2 d$
bits of information with certain probability of success by sending a
qudit.  This is the aim of the present investigation.  The paper is
organized as follows.  First, we illustrate the protocol for exact but
probabilistic super  dense coding for  qubits in section 2. In section
3 we generalize the scheme to higher dimensions. We find that the
success probability of performing super  dense coding is exactly same
as the success probability of distinguishing  a set of non-orthogonal
states.  It is indeed interesting to identify the problem of
probabilistic super  dense coding with  unambiguous state
discrimination. Alternately, one may think that  this problem is
related to unambiguous  discrimination among unitary operators with an
entangled probe state. It has  been shown that a set of unitary
operators can be unambiguously  discriminated iff they are linearly
independent \cite{cm}.  This is true for any  Hilbert space
dimension. Furthermore, any probe state with maximum Schmidt  rank is
sufficient to enable us to do the discrimination. Therefore, we can
say that one   can do probabilistic dense coding with any maximum
Schmidt rank pure  entangled state if you encode the information using
a set of $d^{2}$ linearly  independent unitary operators.  This shows
that the ability to perform super  dense coding  is not only
determined by the amount of entanglement shared between the  sender
and the receiver but also depends on the extent to which the states
encoding the message can be distinguished. In section 4, we
investigate if  the use of more prior entangled state can enhance the
success probability of  performing dense coding. In particular, we
have asked if by sharing a  $(D \times D, D > d)$ entangled state and
by encoding $d^2$ messages in a $D$-state system, one can send
$2\log_2 d$ classical bits in a deterministic fashion? The answer to
this can be negative sometimes.  We find that more entanglement may not be
useful in the sense that  it may or may not enhance the success
probability of performing dense coding. On the contrary, if we use a 
$(D \times D, D > d)$ maximally entangled state and try to send 
$2\log_2 d$ classical
bits, then surprisingly the success probability decreases with
increasing $D$. When $D=d$, then the optimal probability of performing
the dense coding is exactly unity, which is the standard case.
This also helps us to understand why we need $\log_2 d$ ebits to
send $2\log_2 d$ classical bits in a deterministic fsahion.
We end the paper with discussions, conclusions and some future 
directions in section 5.

\section{Probabilistic dense coding with a qubit}

In this section we describe how to send two classical bits $(2 \log_2
2)$ of information in a probabilistic manner using a partially
entangled state.  First we give the most general set of basis vectors
for two qubit Hilbert space. This was introduced in \cite{agrawal} in
the context of  probabilistic teleportation. We can define a set of
mutually orthogonal NME basis vectors $\{ |\psi_i \ra \} (i=1,2,3,4)$
$\in  \cal{H}^{\rm 2} \otimes \cal{H}^{\rm 2}$ as follows
\bea
|\psi_1 \ra &=& |\varphi^{+}_{\ell}\rag  = L \,(|00\rag + \,\ell
\,|11\rag) \nonumber\\ |\psi_2 \ra &=& |\varphi^{-}_{\ell}\rag =  L \,
(\ell^{*}|00\rag - \,|11\rag) \nonumber\\ |\psi_3 \ra &=&
|\psi^{+}_{p}\rag  = P \, (|01\rag + \,p \,|10\rag)  \nonumber\\
|\psi_4 \ra &=& |\psi^{-}_{p}\rag = P \,(p^{*}|01\rag - \,|10\rag)
\eea
Here $\ell$ and $p$ can be complex numbers in general and  $L = \Nol$
and $P = \Nop$ are real numbers.  We notice that when $\ell = p = 0$,
this basis reduces to the computational basis which is not
entangled. For  $\ell = p = 1$, it reduces to the Bell  basis which is
maximally entangled. Therefore this set interpolates between
unentangled and maximally entangled set of basis vectors. Also note
that the set $|\varphi^{\pm}_{\ell}\rag$ and $|\psi^{\pm}_{p}\rag$
have different amount of entanglement for $0 < \ell, p <1$. As
measured by von Neumann  entropy \cite{pr}, the entanglement of
$E(|\varphi^{\pm}_{\ell}\rag)= (- \, L^2\rm{log}_{2}L^2 - L^2 \,
|\ell|^2 \,
\rm{log}_{2}L^2 |\ell|^2)$ and of
$E(|\psi^{\pm}_{p} \rag)= (- \,P^2 \rm{log}_{2}P^2 - P^2 \, |p|^2 \,
\rm{log}_{2}P^2 |p|^2)$, respectively are different for these sets.
However, when $\ell = p$, then all basis vectors have identical von
Neumann entropy. Even though $|\varphi^{\pm}_{\ell}\rag$ and
$|\psi^{\pm}_{p}\rag$ have different amount of entanglement they
satisfy the completeness condition, i.e., $\sum_i |\psi_i \ra \la
\psi_i| =I$ for all $\ell$ and $p$.

For the purpose of super dense coding one may use any one of the NME
basis vectors as a shared resource.  Let Alice and Bob share a
non-maximally entangled state $| \phi_{\ell}^{+}\ra$ as a quantum
channel which is given by
\be
|\phi_{\ell}^{+} \rag = L \,(|00\rag + \, \ell |11\rangle ).
\ee
Here, without loss of generality $\ell$ can be chosen to be a real
number.  Notice that because of the existence of Schmidt decomposition
\cite{ep,akp1} any two qubit entangled state $|\Psi \rag  \in
\cal{H}^{\rm 2} \otimes \cal{H}^{\rm 2}$ such as
\be
|\Psi \rag = a |00\rag +  b |11\rag + c|01 \rag + d |10\rag,
\ee
can be written as a superposition of two basis vectors. In general,
the computational basis states such as $|0\rag$ and $|1\rag$ need not
be the Schmidt basis, but we assume that Alice and Bob know the
Schmidt basis and  coefficients. Then (2) is the most general
non-maximally entangled state up to local unitary transformations
relating Schmidt basis and computational basis states.  By local
unitary transformation, Eqn.(3) can be brought to Eqn.(2).

Let Alice apply on her particle,  any one of the four unitary
operators  $\{I, \sigma_x, i\sigma_y, \sigma_z \}$ that encodes two
bits of classical information. Then, depending on the applied unitary
transformation  the shared state undergoes the following transformation
\bea
|\phi_{\ell}^{+} \rag &\rightarrow&  (I \otimes I)  \phi_{\ell}^{+}
|\rag = |\phi_{\ell}^{+} \rag \nonumber\\ \phi_{\ell}^{+} \rag
|&\rightarrow&   (\sigma_x \otimes I) |\phi_{\ell}^{+} \rag = L (|
|10\ra + \ell |01\ra ) =  {\tilde \psi}_{\ell}^{+} \rag  \nonumber\\
|\phi_{\ell}^{+} \rag &\rightarrow& (i\sigma_y \otimes I)
||\phi_{\ell}^{+} \rag =  L ( -| 10\ra + \ell |01\ra ) =
|\psi_{\ell}^{-} \rag\nonumber\\ \phi_{\ell}^{+} \rag &\rightarrow&
|(\sigma_z \otimes I) |\phi_{\ell}^{+} \rag =  L (| 00\ra - \ell
||11\ra ) = {\tilde \phi}_{\ell}^{-} \rag.
\eea

Now Alice sends her qubit to Bob. Bob has at his disposal two qubits
which  could be in any one of the four possible states $\{
|\phi_{\ell}^{+} \rag,  |{\tilde \psi}_{\ell}^{+}
\rag,|\psi_{\ell}^{-} \rag,  |{\tilde \phi}_{\ell}^{-} \rag \}$. If
Bob is able to distinguish all the four states deterministically then
he can extract two classical bits of  information. However, the above
four states are not mutually orthogonal.  In quantum theory,
non-orthogonal states cannot be distinguished with certainty. Note
that if the shared state is a ME state, then all the above  four
states are mutually orthogonal and the protocol reduces to the
standard one \cite{bw}.

However, it is known that if a set contains non-orthogonal states that
are linearly independent then they can be distinguished with some
probability of success \cite{ivan,peres,chefles,chefles1,duan}.  Now
in our case, it is easy to check that the above  set  $\{
|\phi_{\ell}^{+} \rag,  |{\tilde \psi}_{\ell}^{+}
\rag,|\psi_{\ell}^{-} \rag,  |{\tilde \phi}_{\ell}^{-} \rag \}$ is
actually {\em linearly independent}.  The basic idea is that once Bob
is able to distinguish these states with some probability of success,
then he can know which unitary operation Alice has applied, hence he
can extract two classical bits of information. The optimal probability
of distinguishing these linearly independent states is then the
optimal success probability of performing the super dense coding with
a  partially entangled state.

The way it works is that first Bob performs a projection onto the
subspaces  spanned by the basis states $\{|00\ra, |11\ra \}$ and
$\{|01\ra, |10\ra \}$.  The corresponding projection operators are
$P_1 = |00\ra \la 00| +  |11\ra \la 11 |$ and $P_2 = |01\ra \la 01| +
|10\ra \la 10 |$, where $P_1$ and $P_2$ are mutually orthogonal. If he
projects onto $P_1$, then he  knows that the state is either
$|\phi_{\ell}^{+} \rag$ or $|{\tilde \phi}_{\ell}^{-} \rag$.
Similarly, if he projects onto $P_2$, then he knows that the state is
either  $|{\tilde \psi}_{\ell}^{+} \rag$ or $|\psi_{\ell}^{-}
\rag$. Now the task at Bob's hand is to further distinguish between
these two states within  the given subspace. To achieve this, he
performs a generalized measurement  described by Positive Operator
Valued Measurements (POVMs) on his two  qubit states.  POVMs are
nothing but the generalized measurement operators which can be
realized by enlarging the Hilbert space of the quantum system and
performing orthogonal projections on the ancilla system. They are
described by a set of positive operators $\{ A_{\mu} \}$ that sum to
unity, i.e, $\sum_{\mu}  A_{\mu} = I$. Here, the number of outcomes
can be much larger than the Hilbert space  dimension of the quantum
system, i.e. $\mu \ge d$. Upon measurement, the  probability of
observing ${\mu}$th outcome in a quantum state $\rho$ is given by
$p_{\mu} = {\rm tr}(A_{\mu} \rho)$. In general these POVM's are not
necessarily orthogonal. If they are orthogonal then they reduce to the
standard von Neumann projection operators.

Now the corresponding POVM elements for the two qubit case in the
subspace $\{|00\ra, |11\ra \}$ are given by
\bea
A_1 = \frac{1}{2} \left(\matrix{ \ell^2  & \ell  \cr
\ell  & 1 } \right), ~~
A_2 = \frac{1}{2} \left(\matrix{ \ell^2  & -\ell  \cr  -\ell  & 1 }
\right), ~~ A_3 = \left(\matrix{ 1-\ell^2  & 0  \cr  0  & 0 } \right) .
\eea
This was first given in \cite{ehpp} and also used in conclusive
teleportation
\cite{mor}. One can check that $A_1 + A_2+ A_3 =I$. Here if Bob gets 
$A_1$ then the state is  $|\phi_{\ell}^{+} \rag$, if he gets $A_2$
then it is  $|{\tilde \phi}_{\ell}^{-} \rag$  and if he gets $A_3$
then the result is inconclusive. The success probability of
distinguishing  $|\phi_{\ell}^{+} \rag$ and $|{\tilde \phi}_{\ell}^{-}
\rag$ is $1 - \la \phi_{\ell}^{+}| A_3 |\phi_{\ell}^{+} \ra$ which is
same as $1 - \la {\tilde \phi}_{\ell}^{-}| A_3 |{\tilde
\phi}_{\ell}^{-} \ra$.  This turns out to be equal to $\frac{2
\ell^2}{1 + \ell^2}$. Similarly, for the other two cases one can show
that the success probability is given by the  above expression. Hence,
we can say that Bob can extract two bits of classical information with
a success probability given by $\frac{2 \ell^2}{1 + \ell^2}$. For the
maximally entangled case,  $\ell = 1$ and so probability becomes
one. This is then the standard super  dense coding protocol that works
in a deterministic fashion.  This completes the probabilistic super
dense coding protocol with a qubit.

\section{Probabilistic dense coding for qudit}

We know that if Alice and Bob share a $(d \times d)$ maximally
entangled state then by sending a qudit Alice can communicate $2\log_2
d$ bits of classical  information. Can she send the same amount of
classical information in a  probabilistic manner if they share a
non-maximally entangled state? The answer is yes. Interestingly, this
problem is also directly related to the problem of distinguishing a
set of non-orthogonal states with a certain  probability of success.

In this section we generalize our protocol when Alice and Bob share a
NME state in higher dimensions (say a two-qudit state in $d \times
d$). The shared  NME state is expressed as
\be
|\Psi \ra = \sum_{k=0}^{d-1} \sqrt{ p_k} |k \ra |k\ra,
\ee
where $p_k$'s are the Schmidt coefficients and  $|k\ra$'s are the
Schmidt bases vectors.  Alice and Bob possess one particle each. Now
Alice encodes her $d^2$ possible choices or $2\log_2 d$ bits of
classical information  using unitary operators ${\cal U}_{mn}$, where
$m,n = 0,1, \ldots d-1$.  These unitary operators are given by
\be
{\cal U}_{mn} = (U)^m (V)^n,
\ee
where $U$ is the shift operator and $V$ is the rotation operator whose
action on  the basis states are defined as follows
\bea
U| k \ra &=& |(k \oplus 1) \ra \nonumber\\ V|k \ra &=& e^{2\pi ik/d}
|k\ra
\eea
and $\oplus$ is addition modulo $d$. After Alice applies ${\cal
U}_{mn}$ to her particle the two-qudit state transforms as
\be
|\Psi\ra \rightarrow ({\cal U}_{mn} \otimes I)|\Psi\ra =
\sum_{k=0}^{d-1} \sqrt{ p_k}  e^{ 2 \pi i nk/d} |k \oplus m \ra | k\ra = |\Psi_{mn}\ra.
\ee
Next, Alice sends her qudit to Bob who has the two qudit state
$|\Psi_{mn}\ra$ at his disposal. If Bob is able to perform a
measurement and distinguish all $d^2$ states perfectly then he can
extract $2 \log_2 d$ bits of information deterministically. However,
these $d^2$ states given  above are not orthogonal. Indeed, they
satisfy the following relation
\be
\la \Psi_{mn} | \Psi_{m'n'}\ra = \sum_{k=0}^{d-1} p_k e^{- 2 \pi i k(n-n')/d} 
\delta_{mm'}.
\ee 
Only when all $p_k$'s are same (i.e. the shared state is ME) the above
$d^2$ states are orthogonal. Now Bob has to find a strategy to
distinguish  these states. His ability to distinguish them will decide
the  success or failure to extract $2\log_2 d$ bits of classical
information.  Of course, he cannot do so perfectly. But he can succeed
in distinguishing  the above states with some probability. Then the
probability of  distinguishing  these non-orthogonal states will be
the probability of successful dense coding for a qudit.

Here, we are going to use ideas about discriminating non-orthogonal,
but linearly indepedent quantum states and present a closed form
expression for average success probability of distinguishing a
collection of such quantum states. This is another direction of research
by itself, so we do not intend to review its status here
\cite{ivan,peres,chefles,chefles1}.  Rather we will be using some  of
the results. The pertinent question in the present context is that if
we  have a set that contains a collection of  quantum states $\{
|\Psi_i\ra \} (i=1,2, \ldots N)$ in some Hilbert space,  then can we
perform some measurement and tell in which state the system is? If
these states are orthogonal then the standard von Neumann projection
can give us an answer with certainty.  However, if they are
non-orthogonal then no von Neumann type measurement can unambiguously
identify the states. Then one has to  take recourse to the idea of
generalized or POVM measurements which can help us in discriminating
non-orthogonal states with some probability if and only if the states
are linearly independent \cite{chefles}. A more convenient approach
was suggested by a theorem of Duan-Guo \cite{duan} which tells us that
there is a unitary operator together with post selection of
measurement action which can identify a set of linearly independent
states with some success probability. More precisely it states that
the set  $\{ |\Psi_i\ra \} (i=1,2, \ldots N)$ can be identified,
respectively, with  efficiency $\gamma_i$ if and only if the matrix
$X^{(1)} - \Gamma$ is  positive definite \cite{duan} where $X^{(1)} =
[\la \Psi_i |\Psi_j \ra ]$  is the Gram matrix  and $\Gamma = {\rm
diag}(\gamma_1, \gamma_2, \cdots  \gamma_N)$.  In terms of the unitary
operator on the input and probe state the process takes the form
\bea
U(|\Psi_i \ra |P \ra ) =
\sqrt{\gamma_i} |\Psi_i' \ra |P_i \ra +\sqrt{1- \gamma_i}  
|\Phi_i \ra |P_{N+1}\ra
\eea
where $|P\ra$ is the initial state of the probe, 
$|P_1 \ra, |P_2 \ra, \cdots |P_{N+1} \ra $ are orthonormal basis of the probe 
Hilbert space, $|\Psi_i' \ra$ is the final state of the system, and 
$|\Phi_i \ra$ is the failure component. After the unitary evolution, if we
perform a von Neumann projection on the ancilla system and get 
$|P_i\ra, i =1,2, \ldots N$, 
then we are able to identify the state. But if we get $|P_{N+1}\ra$, then 
we discard it. 
The success probability of identifying these states is $\gamma_i$. 
Using Eqn.(11) we derive the optimal bound
on the success probability of distinguishing any two non-orthogonal but 
linearly indepedent states. Taking the inter inner product we have
\bea
\la \Psi_i | \Psi_j \ra = 
\sqrt{\gamma_i \gamma_j } \la \Psi_i'|\Psi_j' \ra \la P_i |P_j \ra 
+\sqrt{(1- \gamma_i)(1- \gamma_j) }  
\la \Phi_i| \Phi_j \ra.
\eea
Using the above equation we can obtain the tight inequality for distinguishing 
any two non-orthogonal states from the set. It is given by
\bea
\frac{1}{2}(\gamma_i + \gamma_j ) ( 1 - \delta_{ij}) \le 
1 - |\la \Psi_i|\Psi_j \ra|.
\eea
This holds for all $i, j$. For $ i = j$ we have $\gamma_i = 0$.
One may solve a series of inequalities to obtain individual success 
probabilities. However, we are interested in the average success
probability.
This may be obtained as follows. Define the total success probability
as $\gamma = \sum_i \gamma_i$ and the average success probability as
${\bar \gamma} = \frac {\sum_i \gamma_i}{N}$, 
where $N$ is the number of 
linearly independent vectors and $N \le {\rm dim}({\cal H})$. 
Then performing a double sum in the above inequality,
we have the average success probability as
\bea
{\bar \gamma}  \le \frac{N}{N-1} - \frac{1}{N(N-1)} \sum_{i,j=1}^N |\la
\Psi_i|\Psi_j \ra|.
\eea
Alternately, this can be expressed as 
\bea
%\begin{displaymath}
{\bar \gamma}  \le 1 - \frac{1}{N(N-1)} \sum_{i,j=1 \atop i\not=j }^N
|\la \Psi_i|\Psi_j \ra|.
%\end{displaymath}
\eea
This shows that if the set contains states that are orthogonal 
then there is no error, the 
average success probability will be always unity. The second term in the 
optimal success probability represents
the deviation due to the non-orthogonal nature of the states involved. 
To our knowledge such a closed form expression for total or average 
success probability of distinguishing $N$ non-orthogonal states
has not been obtained before. This is another key result of our paper.

Coming back to the super dense coding scheme, once Alice applies $d^2$ 
unitary operators and sends the qudit to Bob, Bob has $d^2$ non-orthogonal 
states $\{ |\Psi_{mn} \ra \}$. The task for Bob is how well he can distinguish 
these states. First, Bob performs $d$ orthogonal projections $P_m = 
\sum_k |k\oplus m\ra \la k \oplus m | \otimes |k \ra \la k|, m = 0, 1, \ldots 
d-1$ that 
projects these states onto $d$ mutually orthogonal subspaces. Now within 
each subspace there are
$d$ non-orthogonal but linearly indepedent states that Bob has to 
distinguish. For example, if
Bob projects onto $P_0$, then this subspace has $\{|\Psi_{0n} \ra \}$ states
which are all non-orthogonal. 
He can perform a unitary operation on two qudits and an ancilla state.
After post selection of measurement outcome (in other words he is
performing a POVM) he can extract $2\log_2 d$ bits 
of information with certain non-zero probability of success. 
The average success probability of
distinguishing $d$ states within a subspace (let us say for $m=0$) can be
obtained from eqn. (14) by putting $N=d$
\be
{\bar \gamma}  \le \frac{d}{(d-1)} - \frac{1}{d(d-1)} \sum_{n, n'=0}^{d-1}
|\la \Psi_{0n}|\Psi_{0n'} \ra| 
\ee
Alternately, the 
average success probability with which he can distinguish $d$ non-orthogonal 
states is given by
\bea
{\bar \gamma}  \le 1 - \frac{1}{d(d-1)} \sum_{n, n'=0 \atop n\not= n'}^{d-1} 
|\sum_{k=0}^{d-1} p_k e^{-2\pi i k (n-n')/d} |
\eea
The protocol works for other subspaces also with the average success 
probability as given in (16). Thus by sharing a partially entangled state
Alice can communicate $2\log_2 d$ classical bits to Bob with a non-zero 
success probability. This completes the super dense coding scheme with any
higher dimensional entangled state.

Just as a consistency check one can also obtain the average success 
probability of performing super dense coding with qubits. 
Recalling from previous section, we
note that after Bob performs projection onto two subspaces he has only two 
non-orthogonal states within each subspace, so $N=2$. Then the 
above relation reduces to
${\bar \gamma} \le 1 - (p_0 - p_1)$. Identifying $p_0 = L^2$ and 
$p_1 = L^2 {\ell}^2 $ we have 
${\bar \gamma} \le
1 - L^2(1 - \ell^2) = 2\ell^2/(1+ \ell^2)$ which was obtained in the 
section 2.
 
As a further illustration of the general result for $d\times d$, 
let us consider probabilistic dense coding for qutrits, i.e., for $d =3$. 
In this case Alice and Bob possess one qutrit each.
%These two qutrits are in a NME state as given by 
%\be
%|\Psi \ra = \sum_{k=0}^{2} \sqrt{ p_k} |k \ra |k\ra.
%\ee
%where as before, $p_k$'s are the Schmidt coefficients and $|k\ra$'s are 
%the Schmidt bases vectors.
Alice can encode $2\log_{2} 3$ bits of information
using unitary operators ${\cal U}_{mn}$, where $m,n = 0,1,2 $.
%on the above state. 
These operators will lead to nine linearly independent states all of which
are not orthogonal.
%These states are the following:
%\be
%|\Psi_{mn}\ra  = \sum_{k=0}^{2} \sqrt{ p_k}  e^{ 2 \pi i nk/3} 
%|k \oplus m \ra | k\ra. 
%\ee
Although these nine states are not mutually orthogonal, they can be
divided into three subspaces, which are mutually orthogonal. The states
in these subspaces are spanned by basis states $\{|00\ra, |11\ra, 
|22\ra\}$, 
$\{|10\ra, |21\ra, |02\ra\}$ and $\{|20\ra, |01\ra, |12\ra\}$
respectively.
By making appropriate Von-Neumann measurements, Bob can distinguish
these three classes. But he cannot perfectly distinguish the states within 
a class, since those states are not orthogonal. However as the states
within a particular class are linearly independent, we can use  
formula (15) to find the probability for Bob to be able to distinguish
these states within a class. This probability
will be the same for all the three subspaces. Let us consider
the states within the class $\{|\Psi_{0n} \ra\}, (n =0, 1, 2)$.
%\bea
%|\Psi_{00}\ra & = & \sqrt{ p_0} |00\ra + \sqrt{ p_1} |11\ra +  
% \sqrt{ p_2} |22\ra, \nonumber \\  
%|\Psi_{01}\ra & = & \sqrt{ p_0} |00\ra + \sqrt{ p_1}  e^{ 2 \pi i/3}|11\ra +  
%                 \sqrt{ p_2}  e^{ 4 \pi i/3}|22\ra, \nonumber \\ 
%|\Psi_{02}\ra & = & \sqrt{ p_0} |00\ra + \sqrt{ p_1}  e^{ 4 \pi i/3}|11\ra +  
%                 \sqrt{ p_2}  e^{ 8 \pi i/3}|22\ra.
%\eea
These states can be distinguished with a success probability
\be
\bar{\gamma} \le 1 - \sqrt{({3\over2} p_{0} - {1\over2})^2 +
                            {3\over4} (p_1 - p_2)^2 }
\ee
Thus, by sharing a $3 \times 3$ NME state Alice can communicate $2\log_2 3$
classical bits with a success probability given in (18). As expected for 
ME
states, $p_0 = p_1 =p_2 = 1/3$ and hence $\bar {\gamma}_{opt} = 1$ which
reduces to the standard case.

\section{Super dense coding with more entanglement}

Since the classical capacity of the communication channel
enhances due to the presence of prior entanglement, one may wonder if the
presence of more entanglement can help to enhance the probability of successful
dense coding when Alice and Bob share a NME.
Specifically, as a result of the above discussion, we ask the question
whether one can send $2\log_2 d$ bits of classical information
by encoding $d^2$ messages in a quDit (a quantum system with 
$D$-dimensional Hilbert space), and sharing a $D\times D$ partially 
entangled state where $D > d$. It may be recalled that recently Gour 
\cite{gour} has investigated the question of teleporting a $d$ level quantum 
system faithfully using a higher dimensional (say $D \times D$ with $D > d$) 
partially entangled state.  He has found that the classical communication
cost of teleporting a qudit is at least $\log_2 (d D)$ bits.

Let the state that Alice and Bob have shared 
is given by
\be
|\Phi \ra = \sum_{\mu =0}^{D-1} \sqrt{ p_{\mu}} |\mu \ra |\mu\ra
\ee
Alice encodes her $d^2$ messages by applying the unitary operators
${\cal U}_{mn}$. Here we have to enlarge the definition of these
operators. The unitary operators ${\cal U}_{mn}$  act as it is given 
earlier by Eqns. (7 - 8) for 
$m,n = 0,1,...,d-1$, while for the rest of the indices, they act as
identity operators. In other words, $U$ acts on the first $d$-dimensional
subspace as defined earlier and as an identity for the rest of the
subspace. After Alice's operation the two-quDit state transforms as 
\be
|\Phi \ra \rightarrow {\cal U}_{mn}|\Phi \ra = |\Phi_{mn} \ra =   
|{\tilde \Psi}_{mn} \ra +  \sum_{\mu =d}^{D-1} \sqrt{ p_{\mu}}
|\mu \ra |\mu \ra
\ee
It can be shown that $|{\tilde \Psi}_{mn} \ra$ are not orthogonal to each
other and also they are not normalized. Similarly the $|\Phi_{mn} \ra$
are non-orthogonal and satisfy
\be
\la \Phi_{mn} | \Phi_{m'n'}\ra = \sum_{k=0}^{d-1} p_k e^{- 2 \pi i k(n-n')/d}
\delta_{mm'} + \sum_{\mu =d}^{D-1} p_{\mu}.
\ee
 
Let us just note that if the shared entangled state 
$|\Phi \ra$ is ME, then $p_{\mu} = 1/D$
and the above orthogonality relation reduces to 
\be
\la \Phi_{mn} | \Phi_{m'n'}\ra = \frac{d}{D} \delta_{mm'} \delta_{nn'} + 
\frac{(D - d)}{D}.
\ee

Now Alice sends her $D$ dimensional particle which encodes her $d^2$ messages.
So basically she has not utilized the total Hilbert space
of her particle. Bob after receiving Alice's particle has the task of 
distinguishing effectively $d^2$ quantum states $\{ |\Phi_{mn} \ra \}$. 
Since the states are not orthogonal, we can conclude that
he cannot discriminate them with certainty and so deterministic dense coding
is not possible. He can however extract $2\log_2 d$ bits of information 
in a probabilistic manner. First, Bob performs the von Neumann projection onto
the $d$ subspaces. Then, he performs POVM's within each subspace to 
distinguish $d$ non-orthogonal states with an average success probability 
\bea
{\bar \gamma}  \le 1 - \frac{1}{d(d-1)} \sum_{n, n' = 0 \atop n\not= n'}^{d-1} 
|\sum_{k=0}^{d-1} p_k e^{-2\pi i k (n-n')/d} + \sum_{\mu =d}^{D-1} p_{\mu} |
\eea
%If we assume that $p_k, k=1 0,1, \ldots d-1$s are same then 
%This probability is clearly smaller than the earlier one given 
%in Eqn.(17) if Alice and Bob have 
%shared a $D \times D$ NME state. For example, if Alice and Bob share a two-qutrit NME state and Alice wishes to
%communicate $2\log_2 2$ classical bits to Bob, then she can do so with a 
%success probability ${\bar \gamma}_{opt} =  1 - (p_0 - p_1 + p_2)$
%which is smaller
%than the success probability ${\bar \gamma}_{opt} = 1 - (p_0 - p_1)$ when
%thay share a two-qubit NME state.
Thus, with a non-zero success probability as given in (23)
Bob can extract $2\log_2 d$ bits of information. But we cannot compare here
whether it is smaller or larger than the previous one (when Alice and Bob
share a $d \times d$ partial entangled state) in general.
%Thus, {\em the use of more entanglement
%does not enhance the success probability of super dense coding}.

The situation is more interesting with maximally entangled states. 
Let us concentrate on the case where Alice and Bob have shared a $D\times D$ 
maximally entangled state. 
Then we have a simple expression for average success probability
which is given by 
\be
{\bar \gamma} \le \frac{d}{D}.
\ee
This shows that if we use a $D \times D$ maximally entangled state and want
to communicate $2\log_2 d$ classical bits then we can do so with an optimal
probability $d/D$. 
This simple expression gives many new insights indeed.
Note that when $d = D$, we have ${\bar \gamma}_{opt} = 1$ 
which is the standard case. However, 
if we use higher dimensional entangled states as shared resource, then the
average success probability is less than one. As we go to higher dimensions
i.e., $D > d$, 
then the average success probability of distinguishing
non orthogonal states decreases.
Thus we can say that the presence of more entanglement in shared states 
may not be always useful (as shown in this scheme). However, it may be too
early to claim that ``the use of more entanglement does not help''.
One could imagine other different schemes where more entanglement may be
useful for dense coding. However, this is not clear at the moment.
For example, Alice could apply $D^2$ unitary operations and
choose $d^2$ of them to encode her messages that are best distinguishable
for Bob. But this is something which needs to be explored in future.
Also, (24) shows that in order to send $2 \log_2 d$
classical bits in a deterministic fashion (i.e., with probability one)
we must have $\log_2 d$ ebits as a shared resource.

We reiterate that the motivation for Section 4 was to see if the mere
presence of more entanglement can enhance the success probability. It
is not the question of using or not using those extra dimensions. (One
may recall that in quantum information processing, sometimes mere
presence of the entanglement can act as a catalyst.) Of course, 
we know that if we use all the entanglement then we
can send more classical bits as entanglement increases in 
determinsitci setting. 
The interesting finding here is that just the mere 
presence of more entanglement may {\em decrease} the probability of the 
success. One would have expected this probability to remain at least the same.

\section{Discussions and Conclusions}

Ideal EPR pairs are very useful for deterministic dense coding, teleportation, 
remote state preparation and so on. However, given arbitrary entangled state
of two qubits or qudits can we do these important tasks. 
In this paper we have investigated the possibility of performing super 
dense coding in a probabilistic manner using a non-maximally entangled 
state as a resource. We have shown 
that $2\log_2 2$ classical bits can be sent probabilistically by sharing 
an entangled state that has less than $\log_2 2$ ebits of entanglement. 
Generalizing to higher dimensions, we have shown that $2\log_2 d$ 
classical bits can be sent in a probabilistic manner using a shared 
entangled state that has less than $\log_2 d$ amount of ebits. The 
success probability of performing super dense coding is related to the optimal
success probability of distinguishing linearly independent 
non-orthogonal states. The expressions for average success probability are 
given for qubit, qutrit as well as for qudit cases.
As far as we know the results presented are new and
do not exist in the literature.

It may be remarked that we could first convert a NME state to a ME
state with certain probability of success \cite{gisin} and then follow
Bennett-Wisner (BW) protocol \cite{bw}. In this scenario the probability of
successful conversion of NME to ME state will be the success
probability of dense coding. However there are two subtle differences
between this case and our protocol. If we first perform local filtering and 
follow BW protocol, the  
probability enters at the stage of conversion of NME state to a ME state. 
Then two classical bits are transferred deterministically. Whereas in our case,
probability enters at a different stage. In our protocol two classical
bits are transfered but recovered probabilistically (due to
non-orthogonal nature of the states). Another important difference is the 
following. In the first case, if we fail in converting NME state to a ME 
state, we cannot use
the Bennett-Wisner protocol. Whereas in our case, whatever is the NME
state, we can send two classical bits to Bob probabilistically. That
is, in our case we will always be able to run the protocol. Therefore, our
protocol is a single shot super dense coding protocol for non-maximally 
entangled resource without first converting to a maximally entangled pair.

In addition to superdense coding for qubit and qudit, we have 
also asked that if one uses a non-maximally entangled 
state in higher dimensions (say $D \times D$), then can one send 
$2\log_2 d, (D > d)$ 
classical bits with a higher probability of success? Interestingly, we find
that the answer to the above question may be negative: more may not be always 
better. But there can be other schemes where entanglement might help.
%We have shown explicitly that if we use more entanglement as a
%shared resource then the success probability of super dense coding decreases. 
We have shown that if we use a maximally 
entangled state in $D \times D$ dimensions, then surprisingly the success 
probability of performing 
super dense coding decreases with increasing $D$. Our analysis also 
explains that to send $2\log_2 d$ classical bits in
a deterministic fashion why one needs exactly $\log_2 d$ ebits and not 
more, not less. 
%Also one may ask if it would be possible to send faithfully 
%$\log_2 (d D)$
%classical bits by sharing a $D \times D$ partially entangled state.

In future it will be interesting to investigate if one can send
$\log_2 (d D)$ classical bits by sharing a $D \times D$ partial entangled
state and sending a qudit. This would be reverse of teleportation
described in \cite{gour}. A priori it does not look like possible, but
it is worth exploring. Also one can see if the probabilistic super
dense coding scheme can be generalized for mixed entangled states. 
That will shed light on the
relation between the classical communication capacity and ability to
distinguish mixed entangled states. It will be of great
value to generalize our protocol for continuous variable quantum systems.
We hope that the probabilistic super dense coding protocol can
be verified experimentally with the present technology.

\vskip 1cm

\noindent
{\bf Acknowledgments:} AKP thanks A. Chefles for useful remarks. 
PP thanks G. Kar for discussions and acknowledges financial assistance 
from DST under the SERC Fast Track Proposal scheme for young scientists.

%\end{multicols}

\end{document}